\documentclass[graphics,floatfix, footinbib,tightenlines,nobibnotes, aps, prl, twocolumn]{revtex4-1}
\usepackage{amsmath,amssymb,wasysym}
\usepackage{graphicx}
\usepackage{dcolumn}
\usepackage{bm}
\usepackage{braket}
\usepackage{subcaption}
\usepackage{verbatim}
\usepackage{float}
\usepackage{bbold}
\usepackage{color}
\usepackage{xcolor}
\usepackage{relsize}
\usepackage{amsthm}
\usepackage{enumerate}
\usepackage{soul,xcolor}
\usepackage[T1]{fontenc}
\usepackage[colorlinks=false ,urlcolor=blue,urlbordercolor={0 1 1}]{hyperref}

\newcommand{\beq}{\begin{eqnarray}}
\newcommand{\eeq}{\end{eqnarray}}

\newcommand{\bsp}{\begin{split}}
\newcommand{\esp}{\end{split}}

\newcommand{\be}{\begin{equation}}
\newcommand{\ee}{\end{equation}}

\begin{document}

\setstcolor{red}

\title{ Twisted Bilayer Graphene Aligned with Hexagonal Boron Nitride: Anomalous Hall Effect and a Lattice Model }
\author{
Ya-Hui Zhang, Dan Mao, and T. Senthil}
\affiliation{Department of Physics, Massachusetts Institute of Technology, Cambridge, MA, USA
}

\date{\today}

\begin{abstract}

A recent experiment reported a large  anomalous Hall effect  in Magic Angle Twisted Bilayer Graphene (TBG) aligned with a hexagonal boron nitride(h-BN) substrate at $\frac{3}{4}$ filling of the conduction band.   In this paper we study this system theoretically, and propose explanations of this observation.  We emphasize that the physics for this new system is qualitatively different from the pure TBG system. The aligned h-BN breaks in-plane two-fold rotation  symmetry and gaps out the Dirac crossings of ordinary TBG. The resulting valence and conduction bands of each valley carry equal and opposite Chern numbers $C=\pm 1$.  A useful framework  is provided by  a lattice extended Hubbard model for this system which we derive. An obvious possible explanation of the anomalous Hall effect is that at $3/4$-filling the system is a spin-valley polarized ferromagnetic insulator where the electrons completely fill a Chern band. We also examine an alternate more radical proposal of a compressible valley polarized but spin unpolarized composite ferm liquid metallic state. We argue that either state is compatible with current experiments, and propose ways to distinguish between them in the future. We also briefly discuss the physics at $1/2$ filling.    

\end{abstract}

\pacs{Valid PACS appear here}
\maketitle

Moir\'e superlattices from twisted Van der Waals heterostructures have emerged as promising platforms to study strongly correlated effects with high tunability\cite{spanton2018observation,cao2018correlated,cao2018unconventional,chen2018gate,yankowitz2018tuning}. Correlated insulators and superconductors have been found  in twisted bilayer graphene and ABC stacked trilayer graphene/hexagonal boron nitride (TG/h-BN) \cite{cao2018correlated,cao2018unconventional,yankowitz2018tuning,chen2018gate}. 

Very recently a  large  anomalous Hall effect was observed \cite{Aaron2019Emergent} in Magic Angle-Twisted Bilayer Graphene (MA TBG) at conduction band filling $\nu=\frac{3}{4}$.  Specifically hysteretic jumps in both the Hall resistivity $(\rho_{xy})$ and longitudinal resistivities $(\rho_{xx})$ were observed. At the lowest temperatures,   $\rho_{xy} \approx 0.5\frac{h}{e^2}$ and $\rho_{xx} \approx 0.3 \frac{h}{e^2}$, corresponding to a large Hall angle, are measured. Evidence for non-local transport, indicative of conducting channels at the sample edge, have been presented.  A key new feature of the device studied in Ref. \onlinecite{Aaron2019Emergent}  is that  one of the graphene layers is nearly aligned with a  hexagonal-Boron Nitride (h-BN) substrate. This alignment has many important effects, as we explain below, and serves to distinguish this system from previous experiments on Magic Angle TBG where no such anomalous Hall effect has been reported. 

In this paper we study theoretically the MA TBG-hBN system, and propose possible explanations of these observations.  A spontaneously spin-valley polarized Chern insulator at $3/4$ filling provides a simple and natural explanation for the large anomalous Hall effect.  We also consider a different novel state which may also explain the data - a compressible composite fermi liquid metal with valley polarization but no spin polarization.  We propose experiments to distinguish these two distinct states. 

In the absence of alignment with h-BN, the moir\'e bandstructure of MA-TBG has ``active" nearly flat bands that are well separated from other bands. The active bands live in each of two Mini Brillouin Zones (MBZ) corresponding to the two valleys of the underlying graphene layers. Within each valley the conduction and valence active bands are connected by Dirac points at the corners of the MBZ. These Dirac points are protected by an excellent emergent $C_2T$ symmetry where $C_2$ refers to a 2-fold rotation and $T$ is time reversal. Either $C_2$ or $T$ maps one valley to the other but their combination preserves the valley index. If however $C_2T$ is broken then the Dirac points will become gapped. Experimentally the presence of Dirac points is evidenced by studying the properties of the system filled to  the Charge Neutrality Point (CNP).  Typically at CNP the system is metallic with a low but non-zero conductance. 

An important effect of alignment with h-BN is that the broken $C_2$ symmetry of h-BN is transmitted to the graphene bands. Thus the Dirac points are gapped and insulating behavior may obtain at CNP.  This is supported by the measured $\rho_{xx}$ at neutrality in Ref.   \onlinecite{Aaron2019Emergent} which is much bigger than the typical measured values in unaligned TBG devices. Furthermore the resulting isolated conduction and valence bands in each valley carry Chern numbers $C = 1, -1$ (opposite valleys carry opposite Chern numbers). Thus the MA TBG-hBN is similar to the many other examples of nearly flat $\pm C$ bands discussed theoretically recently\cite{zhang2018moir}.  As emphasized in Ref. \onlinecite{zhang2018moir}, at total fillings $\nu_T = 1, 3$ nearly flat $\pm $ Chern bands are an excellent platform for the quantum anomalous Hall effect, as well as other even more novel many body states.  Recently Ref. \onlinecite{xie2018nature} described a spin-valley polarized quantum anomalous Hall state in unaligned twisted bilayer graphene where $C_2T$ is broken by interaction effects. 

A further effect of the alignment with h-BN is that there are now two distinct moir\'e superlattices . In addition to the  moir\'e potential produced by the relative twisting of the two graphene layers, the lattice mismatch between h-BN and graphene produces another  moir\'e potential\cite{jung2014ab}. These two  moir\'e lattices have roughly the same period but are rotated by 90$^0$ relative to each other
which makes them  mutually incommensurate. However the strength of the h-BN induced moir\'e potential is expected to be weaker than the other one, and it is a reasonable approximation to ignore it to begin with. It may however play a role by producing in-gap states that may contribute to the lack of exact quan{}tization of the Hall resistivity (in addition to other mechanisms involving disorder)  in the experiments.

The experimental developments on correlated moire superlattices has spawned a large theoretical literature - for a sample  see Refs. \onlinecite{xu2018topological,po2018origin,koshino2018maximally,kang2018symmetry,yuan2018model,dodaro2018phases,ochi2018possible,isobe2018unconventional,padhi2018doped,pizarro2018nature,gonzalez2018kohn,thomson2018triangular,kang2018symmetry,sherkunov2018electronic,huang2018antiferromagnetically,kennes2018strong,liu2018chiral,rademaker2018charge,venderbos2018correlations,guo2018pairing,lin2018kohn,zhang2018low,fidrysiak2018unconventional,roy2018unconventional,su2018pairing,ray2018wannier,zou2018band,po2018faithful,tang2018spin,you2018superconductivity,baskaran2018theory,peltonen2018mean,irkhin2018dirac,wu2018theory,carr2018pressure,guinea2018electrostatic,lian2018landau,lian2018twisted,song2018all,ahn2018failure,laksono2018singlet,chen2018magnetic,liu2019possible}. An important conclusion\cite{po2018origin,zou2018band,song2018all,po2018faithful,ahn2018failure}  is that the bands of TBG have (symmetry) protected topological structure which obstruct the construction of lattice tightbinding models with natural (``on-site") action of all symmetries.  The $C_2T$ breaking induced by the alignment with h-BN however removes this obstruction, and it is possible to construct a lattice tightbinding model to represent the conduction and valence band taken together within each valley.  Unsurprisingly, we show that this takes the form of a lattice Haldane model.  Combining the 2 valleys and projecting the Coulomb interaction yields an effective lattice `extended' Hubbard model suitable for TBG-hBN.  This lattice model provides a useful framework to discuss the physics and may also be useful for future numerical studies.


We consider  twisted bilayer graphene where the top layer is aligned with  the h-BN layer substrate. The twist angle between the two graphene layers $\theta_M$ is chosen to be close to the magic angle $\theta_M=1.05^\circ -1.20^\circ$. The twist angle between the top h-BN layer and the top graphene layer $\theta_{hBN}$ is close to zero.  We assume the bottom h-BN substrate is misaligned and its effect can be ignored.

We use the standard continuum model\cite{bistritzer2011moire} (with $\frac{w_0}{w_1}=0.7$\cite{koshino2018maximally} to account for lattice relaxation)  to calculate the band structure of the TBG/h-BN system.  As time reversal symmetry flips the valley, we can focus only on the band structure within a single valley, say $+$. The Hamiltonian   is
\begin{equation}
  H=H_{TBG}+H_{hBN}
  \label{eq:single_particle_Hamiltonian}
\end{equation}
Here $H_{TBG}$ is the continuum model for the TBG in Ref.~\onlinecite{bistritzer2011moire}. 
 The aligned h-BN has two effects on the top graphene layer:
\begin{equation}
  H_{hBN}=\sum_{\mathbf k}M\psi^\dagger_t \mu_z \psi_t+\sum_{j=1,...,6}\psi^\dagger_t(\mathbf{k+Q'_j})V_j\psi_t(\mathbf{k})
  \label{eq:H_hBN}
\end{equation}
where $\psi_{t,b}$ represent electron destruction operators in the top and bottom valley. 
The first term is an induced   staggered potential on the $A,B$ sublattices of the top graphene layer which acts as a `mass' term.  A rough estimate is obtained from experiments  on monolayer graphene  nearly aligned with h-BN which  show that the band gap at the neutrality point is around $35$ meV. This  implies $M\approx 17$ meV.  The second term in Eq. \ref{eq:H_hBN} represents the moir\'e potential coming from the lattice mismatch between h-BN and graphene. The resulting moir\'e wavevectors are {\em incommensurate} with those associated with the  TBG superlattice. Furthermore a rough estimate from DFT calculations gives $V_j \approx 10$ meV\cite{jung2014ab}  which is much smaller   than  the strength of the TBG moir\'e term (around 110 meV\cite{bistritzer2011moire}), and somewhat smaller than the first term.  Thus as a first approximation we ignore the $V_j$.  This considerably simplifies our analysis as we now have a  well defined band structure in the moir\'e superlattice of TBG.

\begin{figure}[ht]
\centering
\includegraphics[width=0.5\textwidth]{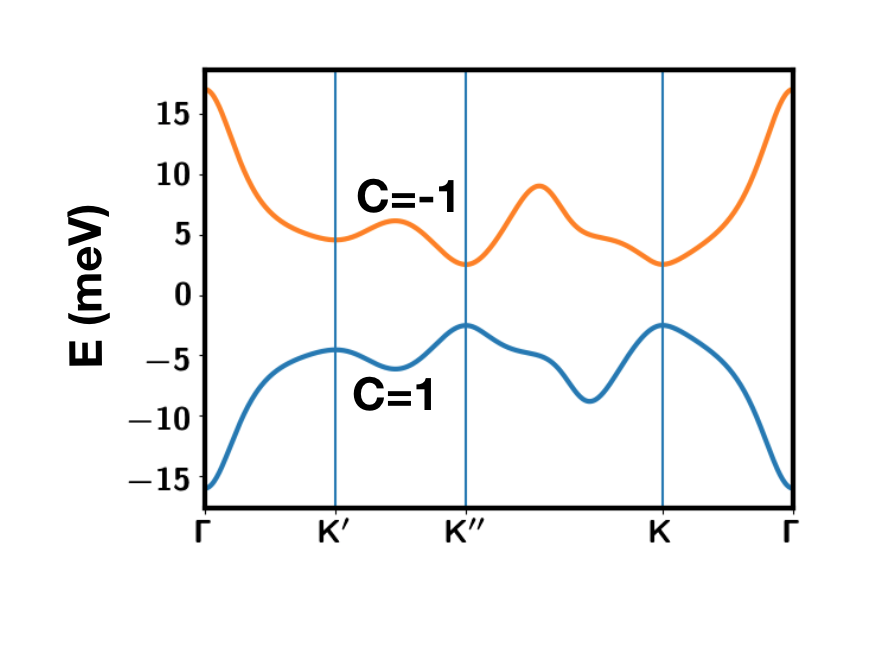}
\caption{Band structure for valley $+$ of the TBG/h-BN system in the MBZ. $\theta_M=1.20^\circ$.   The band of valley $-$ can be generated from the time reversal transformation.}
\label{fig:band}
\end{figure}

The band structure is shown in Fig.~\ref{fig:band} for $M=15$ meV.  As expected, there is a finite band gap around $5$ meV for the value of $M$ we used. Importantly through explicit calculation the conduction and the valence bands for the valley $+$ has Chern number $C=1,-1$.  This Chern number is a simple symptom of the underlying subtle band topology\cite{po2018origin,zou2018band,song2018all,po2018faithful,ahn2018failure} of the unaligned TBG system, and is closely related to the `flipped Haldane model' picture described in Ref. \onlinecite{zou2018band}.

In addition to the Berry curvature, the massive Dirac cone at $K$ or $K'$ point  with gap $\Delta$ have  an out-of-plane orbital magnetic moment proportional to $\frac{1}{\Delta}$\cite{xiao2010berry}. As a result, there is a valley Zeeman coupling $-g_v \mu_B H_z \frac{\tau_z}{2}$ to the magnetic field in the $z$ direction, where $\tau_a,a=x,y,z$ is the Pauli matrix  in valley space.    We numerically calculated $g_v$ for $\theta_M=1.20^\circ$ and $M=15$ meV. Indeed we find that close to the $K$ and $K'$ point, $g_v(K)\approx g_v(K') \approx 15$ for both  conduction and  valence bands, much larger than the spin Zeeman coupling. At neutrality point, this large valley Zeeman coupling can cause splitting between the two valleys in an out of plane magnetic field. This explains  the two fold degeneracy of the Landau fan observed near charge neutrality in Ref. \onlinecite{Aaron2019Emergent}.

Combining the two valleys, we can now project the Coulomb  interaction on to these  bands to obtain an effective model. There are  three important energy scales: the bandwidth $W$ of the conduction band, the band gap $\Delta$ (between conduction and valence bands) and the interaction strength $U$. Our focus is on the experimentally observed correlated insulators at $\nu=1/2$ and $\nu=3/4$ of the conduction band. 

First consider the limit $\Delta \gg U \gg W$. Then  we only need to keep the four conduction bands (including spin and valley)  and the problem reduces to the nearly flat $\pm$ Chern band system studied  in Ref.~\onlinecite{zhang2018moir}.  In the flat band limit, the ground state should be a ferromagnetic insulator from spin or valley polarization. In particular valley polarization is favored over intervalley coherence within a Hartree-Fock calculation\cite{zhang2018moir}.  For $\nu=\frac{3}{4}$, quantum anomalous Hall effect(QAHE) with $|\sigma_{xy}|=\frac{e^2}{h}$ emerges by polarizing both spin and valley. At $\nu=\frac{1}{2}$ in this flat band limit, we expect instead a spin polarized insulator with a quantum valley Hall effect.

Strictly speaking, the TBG/h-BN system is in a different limit $U\sim W>\Delta$ and the detailed many body physics may differ from that discussed in Ref.~\onlinecite{zhang2018moir}. When $\Delta< W$, both the conduction bands and the valence bands should be kept in the  low energy model.   Below  we provide a lattice model by Wannier construction of the active bands which, in contrast to standard TBG,  is possible given the broken $C_2T$ symmetry. The system still has a $C_3$ rotation symmetry.  We take the rotation center as the AA site. The $C_3$ eigenvalues at $\Gamma,K, K'$ are $1,\omega,\omega$ for the conduction band and $1,\omega^*,\omega^*$ for the valence band, with $\omega=e^{i\frac{2\pi}{3}}$. The distinct eigenvalues at $\Gamma, K, K'$ implies that we cannot represent the system on the natural  triangular lattice formed by the AA regions. A honeycomb representation is however possible. The corresponding Wannier functions are readily constructed and have the familiar fidget-spinner shape\cite{po2018origin} reflecting the concentration of charge in the AA regions.

Let  $\mathbf{a_1}=a_M(0,1)$ and $\mathbf{a_2}=a_M(\frac{\sqrt{3}}{2},\frac{1}{2})$ be two basis vectors for the honeycomb lattice, and define the electron operator $c_{i;a\sigma}$ where $a=\pm$ and $\sigma=\uparrow,\downarrow$ are the valley and spin index. The tight binding model takes the form
\begin{equation}
  H_K=-m_0\sum_i (-1)^{X(i)}c^\dagger_{i;a\sigma}c_{i;a\sigma}-\sum_{a\sigma}\sum_{ij}(t^a_{ij} c^\dagger_{i;a\sigma}c_{j;a\sigma}+h.c.)
  \label{eq:H_K}
\end{equation}
where $X(i)=\pm 1$  on the  $A$ and $B$ sublattices. Time reversal symmetry requires that $t^+_{ij}=t^{-*}_{ij}=t_{ij}$.  For each valley this is a modified Haldane model in its topological phase (with a few extra hopping parameters). The tight binding parameters can be found in the supplementary.

\begin{figure}[ht]
\centering
\includegraphics[width=0.5 \textwidth]{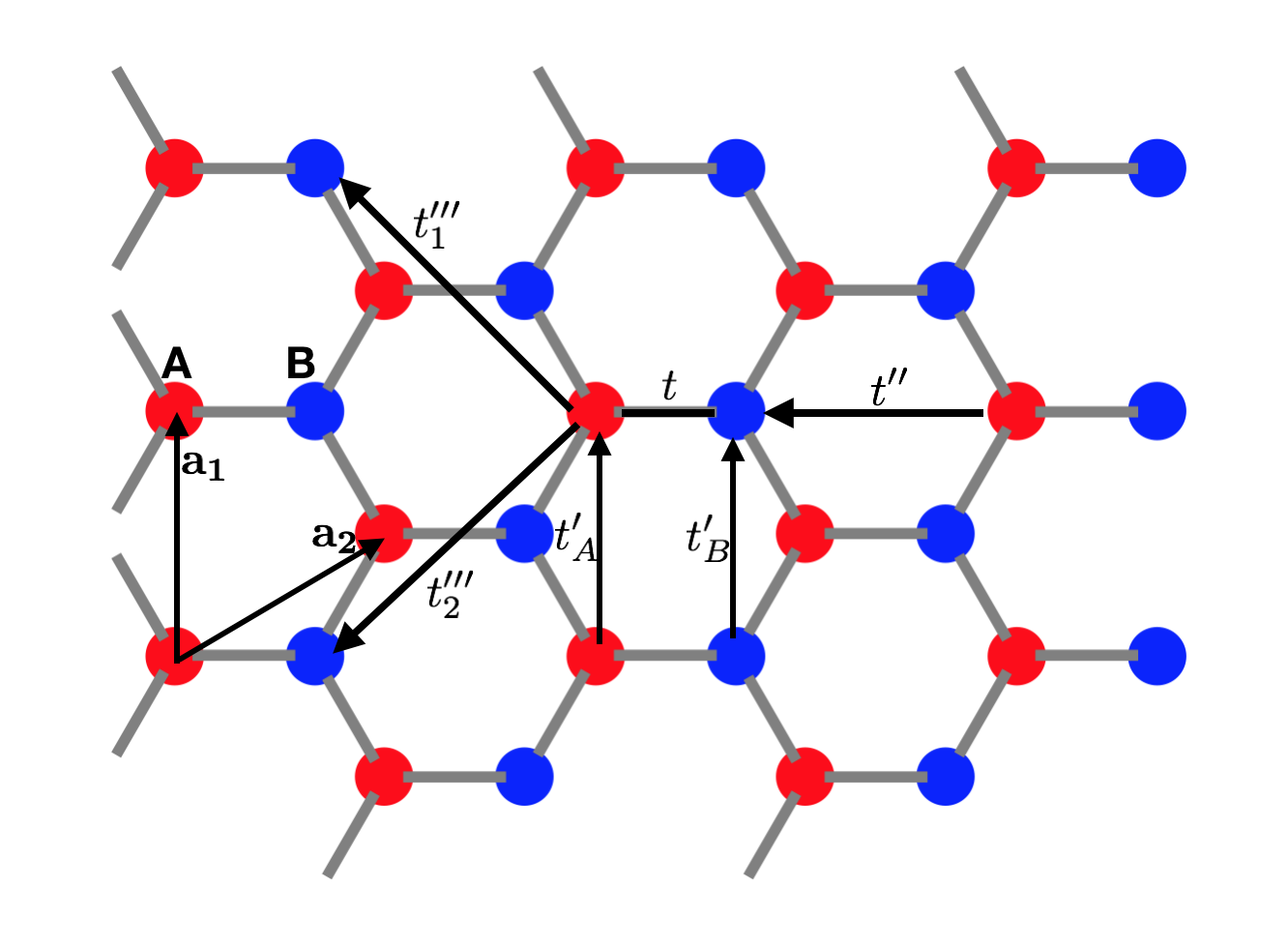}
\caption{Illustration of the lattice model on a Honeycomb lattice.  $C_3 t(\mathbf R) C^{-1}_3=t(C_3 \mathbf R)$  generates inter-sublattice hopping and $C_6 t'(\mathbf R)C^{-1}(6)=t^{'*}(C_6 \mathbf{R})$ generates intra-sublattice hoppings.}
\label{fig:tight_binding_illustration}
\end{figure}

The fidget spinner  Wannier orbital implies that  the interaction is dominated by a cluster charging\cite{po2018origin}  Hubbard interaction.  Furthermore  due to the the non-zero spatial overlap between Wannier orbitals on different sites, there will be an inter-site Hund's term $J$\cite{zhang2018bridging,kang2018strong},    

The interaction term is thus 
\begin{equation}
  H_V=U\sum_{\hexagon}n_{\hexagon}^2-J\sum_{ij}\sum_p S^p_i S^p_j +....
  \label{eq:H_V}
\end{equation}
Here $n_{\hexagon}$ is the electron charge summed over the sites of a hexagonal cluster. $p=1,...,15$ is summed over the $15$ generators of  $SU(4)$, and the ellipses represent other terms (eg a pair hopping) that are less important in the insulator.

Finally we have a lattice model by combining the kinetic and the interaction terms (Eqns. \ref{eq:H_K} and \ref{eq:H_V}). Strictly speaking we also need to add a quasi-periodic potential from the incommensurate h-BN layer(see details in the supplementary).

Given this lattice model, we can consider the strong coupling limit $U \gg \Delta, W$. Then at integer total fillings we will get Mott insulators where the charge on every cluster is frozen.  The corresponding insulator cannot have any Hall response: for a non-zero $\sigma_{xy}$, a Laughlin type of threaded flux induces charge $Q=\sigma_{xy}e$, while in the $U>>\Delta, W$ limit the local density cannot be changed.  Of course the experimental system is likely in the regime $U \sim W \gg \Delta$ and this strong coupling limit is not directly relevant. 

 The observation of a hysteretic anomalous Hall effect  in the experiment\cite{Aaron2019Emergent} clearly shows\footnote{Note that as there is spin $SU(2)$ symmetry, spontaneous spin polarization will not lead to hysteresis; the valley polarization however is an Ising order parameter and hence hysteresis is expected.}  that there is spontaneous time reversal breaking associated with valley polarization  at $\nu=\frac{3}{4}$.  Assuming full valley polarization  we then have a spinful ``Haldane model'' supplemented with interactions at half filling of the conduction band.

If the ground state is fully spin and valley polarized then we get the Chern insulator, and there will be a quantized anomalous Hall effect. In an ideal sample, this state has $\rho_{xy} = \pm \frac{h}{e^2}, \rho_{xx} = 0$ corresponding to a Hall angle of 90$^0$.  Such a spin-valley ferromagnetic insulator state has very good Coulomb energy but has poor kinetic energy. Thus when $W \sim U$ it is interesting to contemplate other states of matter. We will assume full valley polarization in the discussion below. 

For $W \gg U$ a simple Fermi liquid will be the ground state.   This state has $\sigma_{xx} \gg e^2/h, \sigma_{xy} \sim e^2/h$. The Hall conductance is due to a quasiparticle Berry phase that will exist at generic filling of the Chern band. It follows that $\rho_{xy} \ll \rho_{xx}$ so that the anomalous Hall resistivity is small, unlike in the experiments. 

How should we connect the Fermi liquid at $W \gg U$ to the ferromagnetic Chern insulator? A natural possibility (accessed through a Stoner mean field theory) is that the  Fermi liquid first undergoes a transition to a partially spin polarized Fermi liquid which then gives way at larger interaction strengths to the fully spin polarized ferromagnetic insulator. The properties of the partially spin polarized metal will interpolate continuously between those of the spin unpolarized Fermi liquid and the ferromagnetic insulator.  A key experimental signature of this phase will be the presence of two distinct oscillation frequencies (corresponding to the spin split Fermi surfaces) in Shubnikov-DeHaas (SdH) experiments.

We now describe a novel alternate possibility for an intermediate coupling phase. We reason by analogy to a Landau level to which a $C = 1$ band is closely analogous. Since each spin species is at half-filling of the  $C=1$ band,  we may expect the system to be similar to that of spinful electrons in a half-filled Landau level.  In the traditional half-filled Landau level it is well known that a compressible metallic state - the Composite Fermi Liquid  (CFL) - is formed. In the presence of spin it is favorable to instead spin polarize to form a ferromagnetic integer quantum hall state. In contrast to the traditional Landau level, the Chern band has a dispersion. Thus a spin unpolarized Composite Fermi Liquid may be competitive. Such a state  should retain much of the kinetic energy of the simple Fermi liquid while doing better on the Coulomb energy.   A convenient description is through a  parton construction $c_{i;\sigma}=b f_{i;\sigma}$ where the spinless slave boson $b$ carries physical charge and $f_{\sigma}$ is spin-$1/2$ neutral fermion. In the CFL phase,  the boson is at filling $\nu=1$ of a $C = 1$ band and can form  a Fractional Chern insulator (Pfaffian state) 
with $\sigma^b_{xy}=\frac{e^2}{h}, \sigma_{xx} = 0$. Using the Ioffe-Larkin rule\cite{ioffe1989gapless}, we get the resistivity tensor of the physical  electrons:  $\rho^c=\rho^f+\rho^b$. In the clean limit, $\rho^b=\left(\begin{array}{cc}0&-\frac{h}{e^2}\\\frac{h}{e^2}&0\end{array}\right)$ and $|\rho^f|<<1$ is metallic. Then we get $\rho^c_{xy}\sim \frac{h}{e^2}>>\rho^c_{xx}\sim \rho^f_{xx}$. Then the Hall angle $tan^{-1}\left(\frac{\rho^c_{xy}}{\rho^c_{xx}}\right)$  is close to, but strictly smaller, than 90$^0$.  More details  on this CFL phase can be found in the supplementary.

Thus the spin unpolarized composite Fermi liquid provides a concrete  interesting intermediate coupling metallic state with a large Hall angle.    This state will show SdH oscillations with a frequency that, in contrast to the partially spin polarized Fermi liquid,  matches the band theory Fermi liquid.   Other related novel states of matter can also be contemplated but we will leave their elaboration to the future.

Though the Hall angle in the experiments is large, it clearly does not precisely match the expectation of an ideal quantized anomalous Hall system or of the composite fermi liquid.  This   is possibly due both to the presence of disorder and to the presence of the quasi-periodic potential. In particular the quasiperiodic potential may produce nearly extended in-gap states which may   reduce the Hall angle to close to 45$^0$ in the experiment. Finally we remark that  in both the QAH and the CFL state the conduction is predominantly through the sample edge which will lead to  non-local response consistent with experiments.


We now turn to the correlated insulator observed at $\nu=\frac{1}{2}$.  While more exotic phases  cannot be ruled out, the observed two fold degeneracy of the Landau fan\cite{Aaron2019Emergent} suggests a simple picture of a ferromagnetic insulator. Further, based on the experiment result\cite{Aaron2019Emergent} that the resistivity is enhanced with an out of plane magnetic field (which has a valley Zeeman coupling, see the supplementary), we suggest the $\nu=\frac{1}{2}$ insulator has an inter-valley-coherent order with $\tau_{x,y}$ or $\tau_{x,y}\vec{\sigma}$ valley polarization  (see the supplementary). Energetically within a Hartree-Fock theory in momentum space such an IVC order is known to be favored when the anisotropy $\delta \xi(\mathbf k)=|\xi_{+}(\mathbf k)-\xi_{-}(\mathbf k)|$ is large\cite{po2018origin,zhang2018moir,zhang2018bridging}. In the supplementary we show that $\tau_x \vec{\mathbf{\sigma}}$ is selected by the inter-valley Hund's term breaking $U(2)_+\times U(2)_-$ symmetry of separate spin and charge conservation of each valley down to $U(1)_{charge}  \times U(1)_{valley} \times SU(2)_{spin}$. An in-plane magnetic field $H_x$ further favors $\tau_x \sigma_{y,z}$.

In conclusion, we described several aspects of the physics of Magi Angle Twisted Bilayer Graphene aligned with a h-BN substrate.  The $C_2$ breaking due to alignment with h-BN  gaps the Dirac points of TBG, and further renders the conduction and valence bands with Chern numbers $C = \pm 1$. This suggests a natural explanation of the recent observation\cite{Aaron2019Emergent} of a large anomalous Hall effect at $3/4$-filling of the conduction band, as a spontaneously spin-valley polarized ferromagnetic Chern insulator. Energetically such a state is natural when the Coulomb interaction is strong compared to the bandwidth.  At intermediate coupling, other more novel states with a large anomalous but unquantized Hall effect, are possible. The concrete example we discussed - a spin unpolarized composite fermi liquid - may provide an alternate explanation of the data.  We constructed a lattice extended Hubbard model for TBG/h-BN which may be useful for future numerical explorations of intermediate coupling phases. Further this model could also provide an effective model for TBG if $C_2T$ is spontaneously broken.

We thank A. Sharpe and D. Goldhaber-Gordon for sharing their data prior to publication, and for very useful  comments. Our discussion overlaps that of an upcoming paper by N. Bultinck, S. Chatterjee, and M. Zaletel, also on TBG/h-BN. We thank M. Zaletel for inspiring conversations about this and related problems. This work was supported by NSF grant DMR-1608505, and partially through a Simons Investigator Award from the Simons Foundation to Senthil Todadri.

\bibliographystyle{apsrev4-1}
%

\newpage
\widetext \vspace{0.5cm}

\begin{center}
\textbf{\large{}{}{}Supplementary Material for ``Twisted Bilayer Graphene Aligned with Hexagonal Boron Nitride: Anomalous Hall Effect and a Lattice Model{}''}{\large{}{} }
\par\end{center}

\setcounter{equation}{0} \setcounter{figure}{0} \setcounter{subfigure}{0}
\setcounter{table}{0} \makeatletter \global\long\def\theequation{S\arabic{equation}}
 \global\long\def\thefigure{S\arabic{figure}}  \global\long\def\thetable{S\arabic{table}}
\makeatother

\global\long\def\bibnumfmt#1{[S#1]}
 \global\long\def\citenumfont#1{S#1}

\section{Band Structure Calculation}
The continuum model for the TBG is\cite{bistritzer2011moire}
\begin{align}
  H_{TBG}&=\sum_{\alpha=t,b}\sum_{\mathbf k}\psi_\alpha^\dagger(\mathbf k)h^{+}_0 \psi_\alpha(\mathbf k)\notag\\
  &+\sum_{\mathbf k, j=1,2,3}\left(\psi^\dagger_t(\mathbf{k+Q_j}) T_j \psi_b(\mathbf k)+h.c.\right)
  \label{eq:H_TBG}
\end{align}
Here we focus on valley $+$. The other valley is related by time reversal transformation. $\psi_\alpha$ is a two component spinor in terms of $A$ and $B$ sublattice for each layer. $h_0^\dagger$ is the standard Dirac Hamiltonian:
\begin{equation}
  h^+_0=k_x \mu_1+k_y \mu_2
\end{equation}
where $\mu_a$ is the Pauli matrix in the sublattice $A,B$ space for the top layer or the bottom layer. $T_j, j=1,2,3$ is the moir\'e term for the inter-layer coupling. $\mathbf{Q_1}=R_{-\theta_M/2}\mathbf{K_o}-R_{\theta_M/2}\mathbf{K_o}$, where $R_\theta$ rotates a $2D$ vector by $\theta$ around the $z$ direction. $K_o$ is one of the corners of ther original large Brillouin zone.

\begin{equation}
	T_1=w_0-w_1 \mu_1
\end{equation}

$Q_2,Q_3$ and $T_2,T_3$ are generated by $C_3$ symmetry: $\psi_\alpha(\mathbf k) \rightarrow e^{i \frac{2\pi}{3}\mu_3}\psi_\alpha(C_3 \mathbf{k})$.

We use the parameters $w_1=110$ meV and $\frac{w_0}{w_1}=0.7$\cite{koshino2018maximally} to incorporate the lattice relaxation effects. This value of $\frac{w_0}{w_1}$ gives a  hybridization gap around $30-40$ meV between the valence band and the band below for twist angle $\theta_M=1.05^\circ - 1.20^\circ$, consistent with the experimental measurement\cite{cao2018correlated}.

As argued in the main text, the aligned h-BN layer on the top provides a mass term for the top graphene layer:
\begin{equation}
	H_{hBN}=M\sum_{\mathbf k}\psi_t^\dagger(\mathbf k)\mu_z \psi_t(\mathbf k)
	\label{eq:H_hBN_append}
\end{equation}

We then diagonalize $H_{TBG}+H_{hBN}$ from Eq.~\ref{eq:H_TBG} and Eq.~\ref{eq:H_hBN_append} and get the band structure shown in  Fig.1  of the main text. Due to the twist angle $\theta_M$, the original Dirac cone at $K_o$ point for the top graphene layer is put at $K'$ point of the Mini Brillouin Zone (MBZ) while the original Dirac cone at $K_o$ for the bottom graphene layer is put at $K$ point of the MBZ. 

\begin{figure}[ht]
\centering
\includegraphics[width=\textwidth]{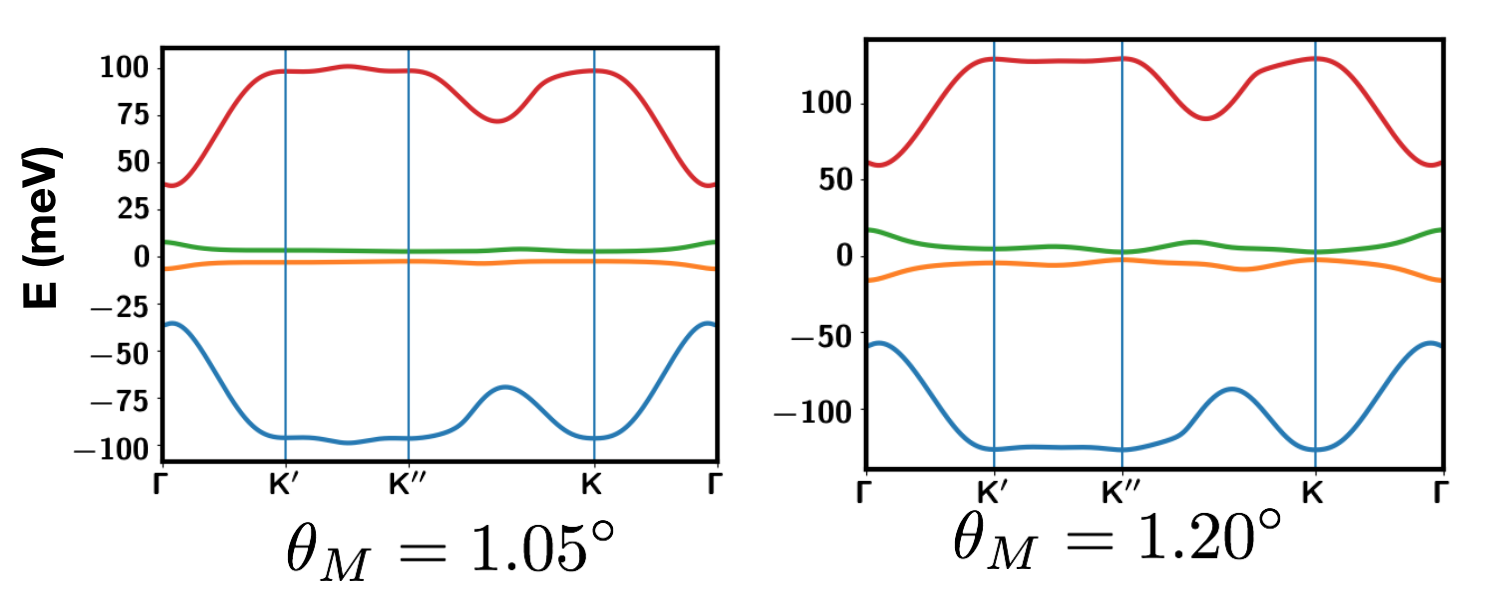}
\caption{Band structure for valley $+$ of the TBG-h-BN system in the MBZ. The middle two bands are well separated from the other bands. The middle two bands have Chern numbers $C=1$ and $C=-1$. We use $\frac{w_0}{w_1}=0.7$ to incorporate the lattice relaxation effects.}
\label{fig:four_bands}
\end{figure}

In the presence of  the $M$ term in the Eq.~\ref{eq:H_hBN_append}, the Dirac crossing is gapped at $K'$ point even if we suppress the inter-layer coupling $T_j$ to zero.  With the inter-layer coupling $T_j$, the Dirac cone at $K$ point is also gapped. The band gap at $K'$ is around $10$ meV while the band gap at $K$ point is only $5$ meV.  Note that there is no symmetry relating $K$ and $K'$.  
Also, the conduction and the valence bands are well separated from the other bands using $\frac{w_0}{w_1}=0.7$, as shown in Fig.~\ref{fig:four_bands}, and from each other. 
Thus the Chern numbers are well defined for both  conduction and valence bands. We calculate the Chern number of each band following  the same method used in Ref.~\onlinecite{zhang2018moir}.  For the valley $+$, we find that the conduction and the valence bands have Chern numbers $C=1$ and $C=-1$ respectively.  Because of the time reversal symmetry, the other valley must have opposite Chern numbers.

\subsection{Valley Zeeman Coupling}
As discussed in the main text, the alignment of the h-BN gives a mass to the Dirac cones at $K$ and $K'$ point in the MBZ for each spin and each valley. As is well known, massive Dirac fermion can have an out of plane magnetic moment, which is opposite for opposite valleys. Therefore there is a valley Zeeman coupling to the out of plane magnetic field:
\begin{equation}
  H_v=- \frac{1}{2}\mu_B H\sum_{\mathbf k} g_v(\mathbf k) c^\dagger(\mathbf k)\tau_z c(\mathbf k)
\end{equation}
with $g_v\sim \frac{1}{\Delta}$. We numerically calculated $g_v$ for $\theta_M=1.20^\circ$ and $M=15$ meV. Indeed we find that close to the $K$ and $K'$ point, $g_v(K)\approx g_v(K') \approx 15$ for both the conduction and  valence bands, which is one order of magnitude larger than the spin Zeeman coupling.  Averaging over the whole MBZ, we have $\bar{g}_v\approx 4$ because the valley Zeeman coupling away from the $K$ and $K'$ points is small.

The valley Zeeman coupling provides a simple explanation of the observed\cite{Aaron2019Emergent} reduction of the degeneracy in the Landau fan emanating from the neutrality point. Close to neutrality the Landau fan is from a Fermi pocket at $K$ point for each flavor (The degeneracy between $K$ and $K'$ is lifted by the alignment of the h-BN layer). The large valley Zeeman splitting (1.5 meV for 1 Tesla) can reduce the degeneracy to only $2$ (coming from spin).

\section{Lattice Model}
The TBG system has a Wannier obstruction to construct a valley preserving and $C_2T$ symmetric model\cite{po2018origin,zou2018band,song2018all,po2018faithful,ahn2018failure}. In the TBG-hBN, the $C_2T$ is broken by the alignment to h-BN. As a result, there is no Wannier obstruction for a valley preserving lattice model. For each valley, the conduction and the valence bands have the opposite Chern numbers, we can build a lattice model by combining both bands.  We constructed the Wannier orbitals following the standard projection methods\cite{marzari2012maximally}. The resulting lattice model is a modified "Haldane model" on a honeycomb lattice for each valley. However, the two lattice sublattice sites correspond to the $AB$ and $BA$ regions, while the density is concentrated on the $AA$ regions. Therefore each Wannier orbital has the shape of a fidget spinner similar to Ref.~\onlinecite{po2018origin}.

\subsection{Tight binding parameters}

In Table.~\ref{table:tight_binding_parameters} we show tight binding parameters defined in the main text for two twist angles.

\begin{table}[H]
\centering
\begin{tabular}{c|c|c|c|c|c|c|c}
$\theta_M$&$m_0$ &$t$ &$t'_A$&$t'_B$&$t''$& $t'''_1$ &$t'''_2$ \\
\hline
$1.08^\circ$ & $1.36$ & $1.22$ &$0.670e^{i0.366\pi}$ &$0.731 e^{-i0.657 \pi}$&$0.801 e^{-i0.685 \pi}$ & $0.123 e^{-i0.48\pi}$&$0.355e^{-i0.411\pi}$\\
\hline
$1.20^\circ$ & $0.076$ & $3.056$ &$0.837e^{i0.56\pi}$ &$0.828 e^{-i0.469\pi}$&$2.062 e^{-i0.54\pi}$&$0.915 e^{-i0.337\pi}$ & $0.815 e^{-i0.434\pi}$\\
\hline
\end{tabular}
\caption{Tight binding model parameters for the modified Haldane model in units of meV.}
\label{table:tight_binding_parameters}
\end{table}

\subsection{Quasi-Periodic Potential}
We also need to add a quasi-periodic potential term from projecting the $V_j$ term in Eq.~\ref{eq:H_hBN} to the two orbitals. 
\begin{equation}
  H_{QP}=V_{QP}\sum_i \cos(\mathbf Q'_j\cdot \mathbf{R_i})c^\dagger_{i;a\sigma}c_{i;a\sigma}
  \label{eq:H_QP}
\end{equation}
where we ignored the quasi-periodicity in the hopping terms for simplicity. The value of $V_{QP}$ can be tuned by displacement field and we keep it as a free parameter.

\subsection{$M\rightarrow 0$ Limit} 
For any finite $M$ we can  build a lattice model on a honeycomb lattice, as done in the main text. This process can be even extrapolated to the $M\rightarrow 0$ limit. How is this consistent with the Wannier obstruction at $M=0$?  In the following we try to resolve this puzzle.  The Wannier obstruction at $M=0$ is related to the $C_2T$ symmetry and is therefore different from the intrinsic Wannier obstruction for the Chern band. Once $C_2T$ breaking is allowed, an exponentially localized Wannier orbital is possible and we can actually recover the gapless Dirac crossing with finite range of hopping, like $R<7$. This process is conceptually similar to the Wannier construction process by ignoring the $U(1)$ valley  symmetry  in Ref.~\onlinecite{po2018origin}.

\begin{figure}[H]
\centering
\includegraphics[width=0.5 \textwidth]{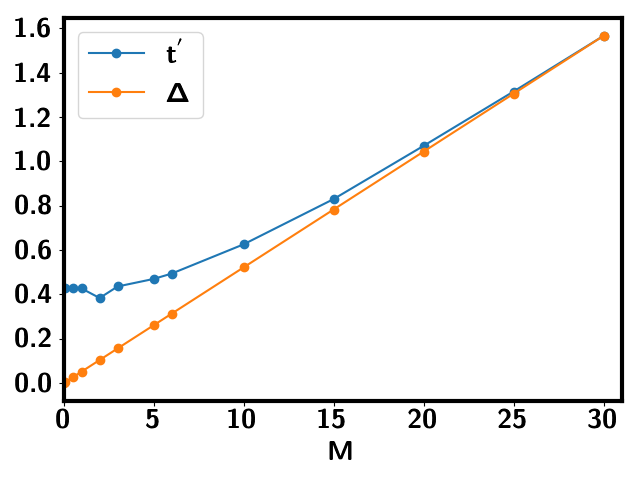}
\caption{The $C_2T$ breaking term in the lattice model $t'$ and the band gap $\Delta$ with $M$. As $M\rightarrow 0$, the lattice model does not have obvious $C_2 T$ symmetry.}
\label{fig:c2t_plot}
\end{figure}

In this process, the lattice model contains $C_2 T$ breaking terms even in the $M\rightarrow 0$ limit.  The typical $C_2 T$ breaking term is the next-nearest neighbor hopping $t'_A\approx -t'_B=t'$. As shown in Fig.~\ref{fig:c2t_plot}, when the external $C_2T$ breaking term $M$ is large, the lattice model has a $C_2 T$ breaking term $t'\propto M$. When $M$ is small, the $C_2T$ breaking term in our lattice model is obviously overestimated. With enough range of hopping, we can still reduce the band gap in the lattice model to be proportional to $M$ at $M\rightarrow 0$ limit. This means that the lattice model with enough range of hopping has a hidden non-local $C_2 T$ symmetry. If one can keep all these non-local  terms in the lattice model, one can still get the correct result of Dirac crossings between valence and conduction bands. However, the purpose of a lattice model is to do a useful approximation. Such a lattice model with a non-local $C_2T$ symmetry(see, eg, Ref.~\cite{kang2018strong}) is dangerous because  it is not clear how to do approximate calculations that maintain this symmetry.  (Similarly, if one does not  insist on a local representation of the  $U(1)_{valley}$ symmetry, a lattice model is also possible in the $M\rightarrow 0$ limit, as explicitly done in Ref. \onlinecite{po2018origin}).     Thus a useful lattice model that keeps just the states from the active bands is  possible only for the large $M$ regime, i.e. when h-BN is aligned. 

\section{Ferromagnetic Insulator at $\nu=\frac{1}{2}$}
At $\nu=\frac{1}{2}$, there is exactly one hole at each honeycomb lattice site. Therefore a strong coupling approach for $U>>t,t'$ with inter-site Hund's term favors a simple insulator which puts a spin-valley polarized hole at each site\cite{zhang2018bridging,kang2018strong}. Either $\tau_z$ or $\tau_x$ valley polarization is selected depending on the competition between the interaction and the kinetic term\cite{zhang2018bridging}. However, this phase leaves both the conduction and  valence band of one flavor empty, which costs kinetic energy.  Therefore, for the experimentally relevant $U\sim W$ regime,  the strong coupling approach is also not appropriate at $\nu=\frac{1}{2}$. 

For $U\sim W$ , a more natural FM order is one where  only the conduction bands of two flavors are pushed up while the other two conduction bands are fully filled. This is described by the  order parameter $c^\dagger \tau_a \sigma_b c $ where $c$ is the electron destruction operator for the conduction band (we suppressed the spin-valley index). $\tau^a$ are Pauli matrices in valley space and $\sigma^a$ are spin Pauli matrices.  

We only consider FM order without momentum dependence ({\em i.e} the particle-hole pair that forms the order parameter has zero internal momentum). There are $15$  such order parameters  corresponding to $\tau^a, \sigma^a, \tau^a \sigma^b$. Our system has an approximate symmetry $U(2)_+\times U(2)_-$\cite{po2018origin,zhang2018moir} generated by $\vec{\sigma},\tau_z \vec{\sigma},\tau_z$ and the total charge.  For any particle-hole order $\psi^\dagger A \psi$ with $A$ a $4\times 4$ matrix, another order is degenerate from the spin-valley rotation: $A\rightarrow U A U^\dagger$ where $U\in U(2)_+\times U(2)_-$. It is then easy to verify that the $15$ FM orders can be grouped to three classes: (1) $\tau_z$; (2)$\tau_z \vec{\sigma},\vec{\sigma}$; (3) $\tau_{x,y}, \tau_{x,y}\vec{\sigma}$.   We can try to decide which of these distinct FM orders is selected by the anisotropies in the Hamiltonian based on a simple Hartree-Fock calculation.  For fully polarized states with $\tau^z$ or $\tau^z \vec \sigma$ ordering, the Hartree-Fock energies are readily seen to be the same. Thus it suffices to compare $\tau^z$ and $\tau^{x,y}$ ordering. As argued in previous papers\cite{zhang2018moir,po2018origin}, within such a Hartree-Fock calculation, either $\tau^z$ or $\tau^{x,y}$ ordering wins depending on the bandwidth. The flat band limit prefers $\tau^z$ while for wider bands it is possible to stabilize $\tau^{x,y}$.  A further selection within each group of orders related by $U(2)_+\times U(2)_-$ occurs through a weak  inter-valley Hund's interaction that locks spins in the two valleys together. 

We therefore proceed phenomenologically, and ask which such order is consistent with results from experiments\cite{Aaron2019Emergent} at $\nu = \frac{1}{2}$. 
First $\tau_z$ can be ruled out because of the absence of anomalous Hall effect at this filling.  Another important information in the experiment is that the resistivity increases with out of plane magnetic field. As discussed in the previous section, the dominant effect of an out of plane magnetic field is a valley Zeeman coupling $-g_v \mu_B H_z \frac{\tau_z}{2}$.   For the $\vec{\sigma}$ and the $\tau_z \vec{\sigma}$ order,   one  valley $+$ and  one valley $-$ band are filled while the other two valley $+$ and $-$ bands are pushed to higher energy with a charge gap $\Delta_c=\Phi-W$ ($\Phi$ is the strength of the splitting of the band due to the order parameter. We assume $\Phi$ is larger than the bandwidth $W$ so as to get an insulator).  With a valley Zeeman coupling, the charge gap $\Delta_c$ decreases: $\Delta_c(H_z)=\Delta_c(0)-g_v \mu_B H$, inconsistent with the experiment.

The only FM order consistent with the experiment is the $\tau_{x,y}$ or $\tau_{x,y}\vec{\sigma}$. They are degenerate  so long as there is  $U(2)_+\times U(2)_-$ symmetry.  However this degeneracy is broken by a weak inter-valley exchange interaction or by an external magnetic field. Let us first consider the former.  The pertinent inter-valley interaction  arises microscopically from large momentum transfer contribution to the density-density repulsion.  Further, for clarity,  we initially present the argument in a form appropriate for a topologically trivial band and later argue that the conclusions remain unchanged for a Chern band - this amounts to initially ignoring form factors associated with Bloch wavefunctions in the interaction.
Therefore we consider 
\begin{equation}
	H_J=\frac{g}{N} \sum_{\sigma_1\sigma_2}\sum_{\mathbf{k_1},\mathbf{k_2},\mathbf{q}}c^\dagger_{+\sigma_1}(\mathbf{k_1+q})c_{-\sigma_1}(\mathbf{k_1})c^\dagger_{-\sigma_2}(\mathbf{k_2-q})c_{+\sigma_2}(\mathbf{k_2})
	\label{eq:inter_valley_Hunds}
\end{equation}
where $g=V(2K_o)$ where $K_o$ is a large momentum in the original large Brillouin zone. $N$ is the number of moir\'e sites. Here $c_{a\sigma}(\mathbf k)$ is the creation operator of the conduction band.   The above term is equivalent to an inter-valley Hund's term \cite{zhang2018moir}:
\begin{equation}
	H_J=- \frac{g}{N} \sum_{\mathbf q} \mathbf{S}_+(\mathbf q) \cdot \mathbf{S}_-(-\mathbf q)-\frac{g}{2N}\sum_{\mathbf q} n_+(\mathbf q)n_-(-\mathbf q)
	\label{eq:explicit_Hund's}
\end{equation}
where $\mathbf{S}_a(\mathbf q)=\sum_{\mathbf k}c^\dagger_a(\mathbf{k+q})\frac{\mathbf \sigma}{2}c_a(\mathbf k)$.
It is more transparent to work   with this form   of Eq.~\ref{eq:inter_valley_Hunds}. 

For simplicity we consider the maximally polarized $\tau_x$ or $\tau_x \sigma_z$ orders. We label S and A as the valley polarization corresponding to $1$ and $-1$ of $\tau_x$.  Then for $\tau_x$ ordering the two filled conduction bands are $S\uparrow, S \downarrow$. For $\tau_x \sigma_z$ ordering, the two filled bands are $S\uparrow, A\downarrow$.   We calculate $\langle H_J \rangle$ using Wick theorem.  From $c_{\pm,\sigma}(\mathbf k)=\frac{1}{\sqrt{2}}(c_{S\sigma}(\mathbf k)\pm c_{A \sigma}(\mathbf k))$ and $\langle c^\dagger_S(\mathbf k)c_A(\mathbf k)\rangle =0$, we have $\langle c^\dagger_{+\sigma}(\mathbf k)c_{+\sigma}(\mathbf k) \rangle=\langle c^\dagger_{-\sigma}(\mathbf k)c_{-\sigma}(\mathbf k)\rangle=\frac{1}{2}\left(\langle c^\dagger_{S\sigma}(\mathbf k)c_{S\sigma}(\mathbf k) \rangle+\langle c^\dagger_{A\sigma}(\mathbf k)c_{A\sigma}(\mathbf k)\rangle\right) $ and $\langle c^\dagger_{+\sigma}(\mathbf k)c_{-\sigma}(\mathbf k) \rangle=\langle c^\dagger_{-\sigma}(\mathbf k)c_{+\sigma}(\mathbf k)\rangle=\frac{1}{2}\left(\langle c^\dagger_{S\sigma}(\mathbf k)c_{S\sigma}(\mathbf k) \rangle-\langle c^\dagger_{A\sigma}(\mathbf k)c_{A\sigma}(\mathbf k)\rangle\right) $.

Labeling $n_{p,\sigma}(\mathbf k)=\frac{1}{2}\langle c^\dagger_{S\sigma}(\mathbf k)c_{S\sigma}(\mathbf k)+c^\dagger_{A\sigma}c_{A\sigma}(\mathbf k)\rangle$ and $n_{m,\sigma}(\mathbf k)=\frac{1}{2}\langle c^\dagger_{S\sigma}(\mathbf k)c_{S\sigma}(\mathbf k)-c^\dagger_{A\sigma}c_{A\sigma}(\mathbf k)\rangle$, we have

\begin{align}
	\langle H_J \rangle&=-\frac{g}{N}\sum_{\sigma}\sum_{\mathbf k,\mathbf q}\langle c^\dagger_{+\sigma}(\mathbf{k+q})c_{+\sigma}(\mathbf{k+q})\rangle \langle c^\dagger_{- \sigma}(\mathbf{k_1})c_{-\sigma}(\mathbf{k})\rangle+\frac{g}{N}\sum_{\sigma_1\sigma_2}\sum_{\mathbf{k_1},\mathbf{k_2}}\langle c^\dagger_{+\sigma_1}(\mathbf{k_1})c_{-\sigma_1}(\mathbf{k_1})\rangle \langle c^\dagger_{-\sigma_2}(\mathbf{k_2})c_{+\sigma_2}(\mathbf{k_2})\rangle \notag\\
	&=-\frac{g}{N} \sum_{\sigma}\sum_{\mathbf{k_1},\mathbf{k_2}}n_{p\sigma}(\mathbf{k_1})n_{p\sigma}(\mathbf{k_2})+\frac{g}{N} \sum_{\sigma_1\sigma_2}\sum_{\mathbf{k_1},\mathbf{k_2}}n_{m\sigma_1}(\mathbf{k_1})n_{m\sigma_2}(\mathbf{k_2}) \notag\\
	&=-\frac{g}{N} \sum_{\sigma}\sum_{\mathbf{k_1},\mathbf{k_2}}n_{p\sigma}(\mathbf{k_1})n_{p\sigma}(\mathbf{k_2})+\frac{g}{N}\sum_{\mathbf{k_1}\mathbf{k_2}}n_m(\mathbf{k_1})n_m(\mathbf{k_2})
	\label{eq:hartree_fock_ivc}
\end{align}
where $n_m(\mathbf k)=n_{m\uparrow}(\mathbf k)+n_{m \downarrow}(\mathbf{k})$ is the difference between  the occupation number of the $S$ band and the $A$ band.

For $\tau_x$ and $\tau_x \sigma_z$ order, the first term is the same: $-2g N$. For the second term, $\tau_x$ order contributes $4g N$ while $\tau_x \sigma_z$ order contributes zero. Therefore we conclude that $\tau_x \vec{\sigma}$ is selected by the term breaking $U(2)_+\times U(2)_-$ down to $U(1)_c\times U(1)_v \times SU(2)_s $.

We now argue the same conclusion holds for a Chern band; then we cannot fix a smooth gauge for $c(\mathbf k)$. Instead, we need to use $c_{a\sigma}(\mathbf k)\rightarrow e^{-i\theta_{a}(\mathbf k)}c_{a\sigma}(\mathbf k)$ in the interaction in Eq.~\ref{eq:inter_valley_Hunds}. In the Hatree-Fock calculation, these additional form factors $e^{-i\theta_a(\mathbf k)}$ cancel in the first term in Eq.~\ref{eq:hartree_fock_ivc}. The form factors need to enter in the second term. However, we also need to add the same form factors in the eigenstate of $\tau_x$.  We have $c_{\pm;\sigma}(\mathbf k)=e^{i\theta_{\pm}(\mathbf k)}\frac{1}{\sqrt{2}}(c_{S\sigma}(\mathbf k)\pm c_{A\sigma}(\mathbf k))$. Then the form factor $e^{-i\theta_a(\mathbf k)}$ will also be cancelled in the second term of Eq.~\ref{eq:hartree_fock_ivc} and we can express it in terms of the gauge invariant term $c^\dagger_S(\mathbf k)c_S(\mathbf k)$ and $c^\dagger_{A}(\mathbf k)c_A(\mathbf k)$. Finally we reach the conclusion that for the Chern band, $\tau_x \vec{\sigma}$ is also selected.

In the above calculation, we assume maximally polarized $\tau_x$ or $\tau_x \vec{\sigma}$ order.  In reality the IVC order may not be maximally polarized. While we do not have a proof for  this more  complicated case, the above analysis suggests that $\tau_x \vec{\sigma}$ order may also be selected.

In the presence of an out of plane  magnetic  field, the mean field Hamiltonian (assuming $\tau_x \sigma_z$ order)   is:
\begin{equation}
   H_M=- \sum_{\mathbf k}c^\dagger(\mathbf k) \left(\frac{\Phi}{2}\tau_x \sigma_z+\frac{1}{2} g_v(\mathbf k)\mu_B H_z \tau_z \right) \psi_c(\mathbf k)
\end{equation}

Because $\tau_x$ anti-commutes with $\tau_z$, the  charge gap in the insulator is   $\Delta_c(H_z)=\sqrt{\Phi^2+g_v^2\mu_B^2H_z^2}-W$.  If, as is reasonable,  $\Phi$ has only a weak dependence on $H_z$, we conclude that the charge gap $\Delta_c$ can be greatly enhanced by the valley Zeeman field given that $g_v$ is large(around $15$ close to the $K$ point), in agreement with what is measured.  This conclusion does not depend on the detailed selection between $\tau_x$ and $\tau_x \vec \sigma$ discussed above.

The response to the in-plane magnetic field $H_x$ (assuming it couples predominantly to the spin) however depends on whether $\tau_x$ or $\tau_x\vec{\sigma}$ is selected.  
 Such a field can further split the energy degeneracy among $\tau_x \vec{\sigma}$. Because there is no spin magnetization, the field energy at first order of perturbation  vanishes. For the second order perturbation, $\tau_x \sigma_{y,z}$ order can have a negative energy correction because $\tau_x \sigma_{y,z}$ anti-commutes with $\sigma_x$. Therefore in-plane magnetic field favors $\tau_x \sigma_{y,z}$. The splitting is of order $\frac{g_s^2 \mu_B^2H_x^2}{\Phi}$ and therefore is small.

\section{Spinful Composite Fermion Liquid}
We give theoretical descriptions of several spinful CFL phases for the filling $\nu_T=\frac{1}{2}+\frac{1}{2}$ of the spinful $C=1$  Chern band.  As discussed in the main text, in the strict flat band limit, the simple ferromagnetic insulator will win. But the states discussed in this Appendix may be competetive once band dispersion becomes significant, {\i.e}, for intermediate coupling $U \sim W$.

We do a  slave-boson parton construction $c_{i;\sigma}=b f_{i;\sigma}$ (We can also do a slave fermion parton, which leads to a "quantum Hall spin liquid" insulator mentioned briefly at the end of this section.). $b$ is a spinless boson which carries the physical charge while $f_\sigma$ is a neutral spin-$1/2$ fermion. We have filling $n_b=1$ and $\sum_{\sigma}n_{f;\sigma}=1$. Besides, $b$ and $f$ need to couple to an internal $U(1)$ gauge field $a$ with opposite charges. 

In this parton construction we can access different phases by putting $b$ and $f$ in different phases. For the fermion $f$, the most natural ansatz is just a spin unpolarized state that with a fermi surface for each spin component.  The spinless boson at $\nu=1$ of a $C=1$ Chern band can be either a Pfaffian state or itself form a Composite Fermi Liquid phase.  For simplicity and because it is somewhat more familiar, here we focus on the former case. Then the boson has a quantum Hall effect with  $\sigma_{xy}^b=\frac{e^2}{h}$.  Such a phase has Ising anyons and the low energy effective theory is denoted $(U(1)_4\times Ising) / Z_2)$. In our case we need to further couple $b$ to the internal gauge field $a$. For the purpose of the charge response of the microscopic electron, we can ignore the non-abelian Ising part and just write down the response of the slave boson $b$ to the gauge field $A-a$ it couples to ($A$ is the external probe electromagnetic gauge field).  This is just a Chern-Simons term $\frac{1}{4\pi}(A-a)d(A-a)$. The low energy theory for the microscopic electron $c$ is
\begin{equation}
	L=\sum_\sigma L[f_\sigma,a]+\frac{1}{4\pi}ada-\frac{1}{2\pi}A da+\frac{1}{4\pi}AdA+...
	\label{eq:spinful_CFL_2}
\end{equation}

 This action  resembles that of the  standard Halperin-Lee-Read theory for the half filled Landau level\cite{halperin1993theory}. However, for this state other terms need to be included to describe the Ising anyon of the slave boson though we will not explicitly write them here.  For discussing low energy electrical transport  the action above which describes the Fermi surfaces and the gauge field $a$ is sufficient. In this sense, this  CFL phase should have essentially the same properties as the conventional CFL phase.  Close to the edge however, a neutral majorana mode may  be present unlike the conventional composite fermi liquid. 

 From the Ioffe-Larkin rule\cite{ioffe1989gapless}, the resistivity tensor of the original electron is
\begin{equation}
  \rho^c=\rho^f+\rho^b
\end{equation}

Therefore we have:
\begin{equation}
  \rho^c=\rho^f+\left(\begin{array}{cc} 0&-\frac{h}{e^2}\\\frac{h}{e^2}&0\end{array}\right)
\end{equation}

In the clean limit, $\rho^f$ behaves like a metal and thus $|\rho^f|<<\frac{h}{e^2}$.  Therefore $\rho^c_{xy}\approx \frac{h}{e^2}>>\rho^c_{xx}=\rho^f_{xx}$. Thus this phase has a large Hall angle, together with non-zero bulk dissipation.

Finally, we point out that the above CFL phase can go through a continuous phase transition by pairing of the composite fermions. In the simplest case, we just consider a spin singlet pairing $\langle f^\dagger_{\uparrow}f^\dagger_{\downarrow}\rangle \neq 0$, the resulting phase is an insulator with Hall conductivity $\sigma_{xy}=\frac{e^2}{h}$.  The charge response is actually the same as the spin polarized Chern insulator. However, in this insulator the spin is in a singlet phase, and the elementary spin excitations  are gapped spinons carrying spin $1/2$, just like a $Z_2$ spin liquid. We dub this exotic insulator as "quantum Hall spin liquid".  It is a   non-trivial non-Abelian topological ordered phase. For example, the  "vison" excitation in a conventional $Z_2$ spin liquid now carries $1/2$ charge and is an Ising anyon, though it still has $\pi$ mutual statistics with the gapped spinon.  Details of this and other "quantum Hall spin liquid"  phases will be discussed elsewhere.

\bibliographystyle{apsrev4-1}
\bibliography{TBG}
\end{document}